\documentclass[%
 reprint,
 amsmath,amssymb,
 aps,
 prstab,
 floatfix,
]{revtex4-1}

\usepackage{graphicx}					% Graphics
\usepackage{wasysym}					% Bold PI
\usepackage{mathtools}					% Curved arrow
\usepackage[makeroom]{cancel}				% Crossed math

\newcommand{\barr}{\begin{array}}
\newcommand{\earr}{\end{array}}

\newcommand{\beq}{\begin{equation}}
\newcommand{\eeq}{\end{equation}}

\newcommand{\beqa}{\begin{eqnarray}}
\newcommand{\eeqa}{\end{eqnarray}}

\begin{document}

%===============================================================================
\begin{abstract}
A problem of coupled-beam instability is solved for two multibunch beams with slightly different revolution frequencies, as in the Fermilab Recycler Ring (RR). Sharing of the inter-bunch growth rates between the intra-bunch modes is described. The general analysis is applied to the RR; possibilities to stabilize the beams by means of chromaticity and feedback are considered.   
\end{abstract}
%===============================================================================

%===============================================================================
\title{ Coupled-Beam and Coupled-Bunch Instabilities}
\author{A.~Burov}
\email{burov@fnal.gov}
\affiliation{Fermilab, PO Box 500, Batavia, IL 60510-5011}
\date{\today}
\maketitle
%\tableofcontents
%===============================================================================

%==============================================================================%
%==============================================================================%
%==============================================================================%

\section{Introduction}

\begin{figure}[b!]
\includegraphics[width=0.95\linewidth]{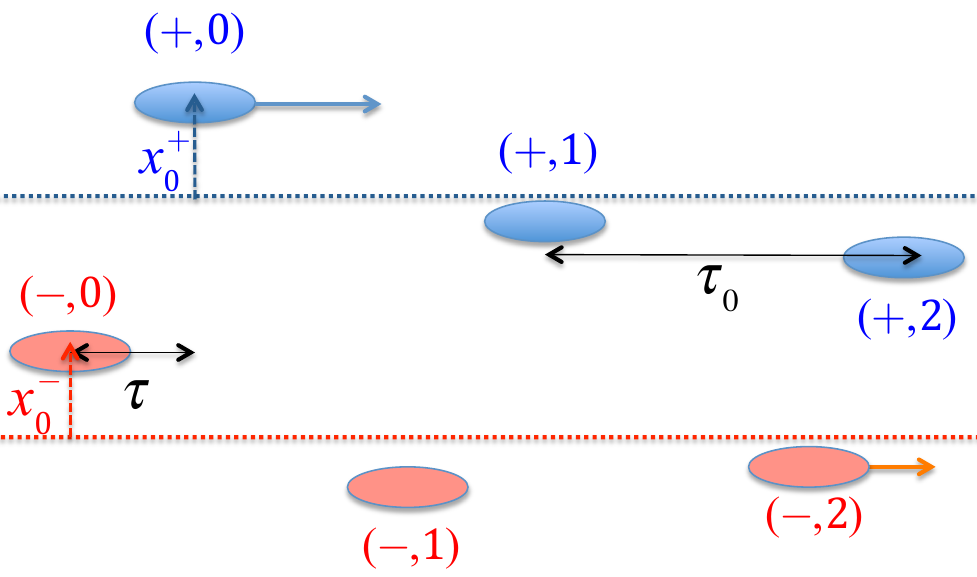}
\caption{\label{fig:SlippingBeams}
Two slipping multibunch beams in the Recycler Ring, when one with a slightly higher energy (blue, marked with $+$) slowly outruns another (red, marked with $-$). Bunch numbers are indicated as $(\pm,\,n)$; the transverse offsets of $(+,0)$ and $(-,0)$ bunches, $x_0^\pm$, are shown.}
\end{figure}
Slip-stacking is a method to increase beam intensity in a synchrotron by merging two beams. When the first beam is moving along its orbit, the second one is injected at a slightly different energy, so that it slips along the first beam while its RF buckets are gradually filled. As soon as that injection is over, when bunches of the first and second beams line up with each other, a sufficiently high RF voltage captures these bunch pairs in the same buckets. Schematically, such slipping motion is shown in Fig.\ref{fig:SlippingBeams}. This method is successfully realized in the RR \cite{Kourbanis:2014iza, Kourbanis:2017zqt}, with about 500 bunches in each beam. Such a high number of bunches makes coupled-bunch interaction a powerful source of collective instabilities. However, the relative motion of the two beams prevents a straightforward application of the existing theory, and requires certain modifications. This paper presents a general solution of this coupled-beam problem as well as some details about the distribution of the inter-bunch tune shifts between the intra-bunch modes, with an application of all that to the Recycler Ring.

\section{\label{Sec:MainEq} Main Equations}

To begin, let us consider every bunch as a macroparticle, with a transverse dynamic offset of $k$-th bunch of the faster beam $x_k^+$, and the same thing for the slower beam $x_k^-$. The equations of motion for these offsets can be presented as follows:  

\beqa 
&& \ddot{x_k^{+}}(t) + \omega_b^2 x_k^{+}(t)=F_k^{++}(t) + F_k^{+-}(t) \nonumber \\ 
&& \ddot{x_k^{-}}(t) + \omega_b^2 x_k^{-}(t)=F_k^{-+}(t) + F_k^{--}(t);  \nonumber \\ 
&& F_k^{++}(t) = 2\omega_b \sum_{n=1}^{\infty}W(n\tau_0) x_{k+n}^{+}(t-n\tau_0) ; \nonumber \\
&& F_k^{+-}(t) = 2\omega_b  \sum_{n=1}^{\infty}W(n\tau_0 - \tau) x_{k+n}^{-}(t-n\tau_0+\tau);  \label{MainEq} \\
&& F_k^{-+}(t) =  2\omega_b \sum_{n=0}^{\infty}W(n\tau_0 + \tau) x_{k+n}^{+}(t-n\tau_0-\tau);  \nonumber \\
&& F_k^{--}(t) = 2\omega_b \sum_{n=1}^{\infty}W(n\tau_0) x_{k+n}^{-}(t-n\tau_0); \nonumber \\
&& x_{k+M}^{\pm}=x_{k}^{\pm}; \ \  \tau =\tau_0 t/T . \nonumber
\eeqa
Here $t$ is time, $\omega_b=Q_x\, \omega_0$ is the betatron frequency, with $\omega_0$ as the revolution frequency and $Q_x$ as the betatron tune, while meanings of bunch separations $\tau$ and $\tau_0$ should be clear from Fig.\ref{fig:SlippingBeams}; $T$ is time required for the slippage per bucket (i.e. per $\tau_0$), and $M$ is the number of bunches per beam, so the total number of bunches in the machine is $2M$. The normalized wake function $W$ is a product of the conventional dipole wake function of the whole ring $W_\perp$ (see e.g. Ref~\cite{chao1993physics}) and the intensity coefficient $N_b r_0 c/(4\pi \gamma Q_x)$, where $N_b$ is the number of particles per bunch, $r_0$ is the classical radius, $c$ is the speed of light, and $\gamma$ is the relativistic factor. With the slip factor $\eta=-(p/\omega_0) d\omega_0/dp$ and the beam-beam relative momentum separation $\delta p/p$, the revolution time $T_0=2\pi/\omega_0$, the slippage period 
\beq
T=-\frac{T_0}{M\eta (\delta p/p)}.
\label{SlipTime}
\eeq

Due to spacial periodicity, the offsets can be expanded over Fourier components, so that for each spacial coupled-bunch harmonic $x_k^{\pm}(t)=x_0^{\pm}(t) \exp(i \phi k)$, with $\phi=2\pi \mu/M$, where the mode numbers $\mu$ are $M$ consecutive integers starting at an arbitrary one. The next step is transition to slow amplitudes, which is conventionally done by the substitution $x_0^{\pm}(t)=a^{\pm}(t) \exp(-i \omega_b t)$. However, this representation of the coupled-beam oscillations is not quite satisfactory yet for slipping beams, since it associates mode amplitudes with specific (zero) bunches. It is important that these bunches do not remain at the same distance from each other. If at $t=0$ the bunches $(+,0)$ and $(-,0)$ exactly align, after time $T$ the bunch $(+,0)$ aligns with the bunch $(-,1)$, while the bunch $(-,0)$ aligns with $(+,-1)$. Thus, the association of the mode amplitudes with specific reference bunches is not adequate to the problem of coupled oscillations of slipping beams. Two-beam collective motion requires such amplitudes that a time shift by $t=T$, when $\tau \rightarrow \tau+\tau_0$, would not change relative phases of the neighbor $+$ and $-$ bunches. This goal is achieved with the following modified amplitudes:       
\beq
b^{\pm} = a^{\pm} \exp(\mp i \phi \tau/(2\tau_0)). \label{atob}
\eeq
For the amplitudes $b^\pm$, the relative phases of the aligned bunches are fully determined by the relative phases of the amplitudes, without any time-dependent explicit factors. For instance, at $t=T$, when $\tau=\tau_0$, the offset of the bunch $(-,0)$ is $x_0^-(T)=b^-e^{-i \phi /2}$ (the common factor $e^{-i\omega_b T}$ is omitted). At that moment, the bunch is aligned with the one numbered $(+,-1)$, which offset is $x_{-1}^+(T)=b^+e^{i \phi /2} e^{-i \phi}=b^+e^{-i \phi/2}$, so their relative phases are equal to those of $b^-$ and $b^+$, as it is the case for any aligned pair of bunches any time the bunches are aligned. 

In terms of the slipping-beam amplitudes $b^{\pm}$, the equations of motion can be written,
\beqa 
&&T \dot{b}^{+} =-i\frac{\phi}{2} b^{+} + i \mathsf{S} b^{+} +i \mathsf{O}(\tau_0-\tau) b^{-};   \nonumber \\ 
&&T \dot{b}^{-} = i\frac{\phi}{2} b^{-}+ i \mathsf{S} b^{-} +i \mathsf{O}(\tau) b^{+};   \label{MainEqFourierb} \\ 
&& \mathsf{O}(\tau)= \sum_{n=0}^{\infty}W(n\tau_0 + \tau) \exp(i \psi(n + \tau/\tau_0)); \nonumber \\
&& \mathsf{S}=\mathsf{O}(\tau_0) ; \ \ \     \psi = \phi+\omega_b \tau_0 .  \nonumber  
\eeqa
Here, the wake Fourier series $\mathsf{S}$ and $\mathsf{O}$ describe actions of the {\it{same}} ($\mathsf{S}$) and {\it{other}} ($\mathsf{O}$) beams. For a given wake function, they can be computed and tabulated as certain functions of the coupled-bunch mode, represented by the mode parameter $\psi$, and the slippage phase $\tilde{\tau}=\tau/\tau_0$,  $0<\tilde{\tau} \leq 1$. Without loss of generality, the mode phase $\psi$ can be chosen so that $|\psi| \leq \pi$, the convention held below. When the beam-beam interaction is suppressed for one or another reason, the {\it other} term $\mathsf{O}$ can be dropped, and the well-known single-beam coupled-bunch formulas can be obtained.

Let us now take one more step and include an important parameter, unaccounted for as of yet, chromaticity $\xi=p\,dQ_x/dp$. With time measured in the units of the slipping period $T$, this yields,
\beqa 
&& \dot{b}^{+} =-i\frac{\phi+\chi}{2} b^{+} + i \mathsf{S} b^{+} +i \mathsf{O}(\tau_0-\tau) b^{-};   \nonumber \\ 
&& \dot{b}^{-} = i\frac{\phi+\chi}{2} b^{-}+ i \mathsf{S} b^{-} +i \mathsf{O}(\tau) b^{+},   
\label{MainEqFourierbChi} 
\eeqa
where the chromatic beam-beam phase
\beq
\chi=-\frac{\xi}{\eta}\,\frac{2\pi}{M}
\label{chibb}
\eeq
is the chromatic frequency shift $\omega_0\, \xi \,\delta p/p$ in the units of the inverse slippage period $T$. 

Before going into details of the general solution of Eqs.~(\ref{MainEqFourierbChi}), it would be reasonable to solve them for an important case when the bunch-to-bunch phase $\psi $ is so small that the beam-beam interaction function $\mathsf{O}(\tau)$ can be taken as constant, $\mathsf{O}(\tau) = \mathsf{S}$, which allows to treat the beams as coasting. Substituting $b^\pm \propto \exp(-i\omega t)$, the two eigenfrequencies are obtained, 
\beq
\omega = -\mathsf{S} \pm \sqrt{\mathsf{S}^2+\,(\phi+\chi)^2/4}.
\label{CoastFreq}
\eeq
which can also be found from Eq.~(6.258) of Ref.~\cite{chao1993physics}, assuming the beam longitudinal distribution to consist of two delta-functions. This solution shows that there are two extreme situations with respect to the beam-beam interaction. If the slip phase is small in comparison with the interaction function, $|\phi+\chi|/2 \ll |\mathsf{S}|$, the two beams are either in phase, with the common mode frequency $\omega \approx -2 \mathsf{S}$, or out of phase, when their wakes almost cancel each other out. In the opposite situation of a large slip phase, $|\phi+\chi|/2 \gg |\mathsf{S}|$, the beams essentially do not interact; each of them oscillates with its own frequency $\omega = -\mathsf{S} \pm (\phi+\chi)/2$. Due to wake properties, the self-interaction function $\mathsf{S}(\psi)$ corresponds to instability, $\Im \mathsf{S}(\psi)<0$, only if its argument $-\pi<\psi<0$. At first glance, one may conclude from here that the maximally effective suppression of the instability by the chromaticity requires the conventional rule for the chromaticity sign to be obeyed: the sign of the chromatic phase $\chi$ has to be negative, i.e. the sign of the chromaticity $\xi$ has to be negative below transition and positive above. As it will be seen below in this paper, the situation is, in fact, more complicated.    

Let us come back now to the general case of arbitrary bunch-to-bunch phase $\psi$, Eq.~(\ref{MainEqFourierbChi}). This pair of linear ordinary differential equations can be further simplified with the substitution $b^{\pm}=c^{\pm} e^{i\mathsf{S}}$, which eliminates the time-independent same-beam factor $\mathsf{S}$: %%
\beqa 
&& \dot{c}^{+} =-i (\psi+\Delta \psi) c^{+}/2  + i \mathsf{O}(\tau_0-\tau) c^{-};  \label{MainEqFourierc}  \\ 
&& \dot{c}^{-} = i (\psi+\Delta \psi) c^{-}/2 + i \mathsf{O}(\tau) c^{+},   \nonumber 
\eeqa
where $\Delta \psi = \chi - \omega_b \tau_0$ can be termed the beam-beam phase shift. 
Thereby, the problem is reduced to the pair of ordinary linear homogeneous differential equations with time-dependent coefficients. Its periodical map $\mathcal{P}$ can be obtained by numerical integration: 
\beq
\mathbf{c}(1)=\mathcal{P} \mathbf{c}(0); \ \mathbf{c}=(c^{+},c^{-})^T  \label{Mapc}.
\eeq
Slipping-beam collective modes are described by the eigensystem of the matrix $\mathcal{P}$. Its eigenvalues $\lambda_{1,2}$ give the growth rates $r_{1,2}$ and phase shifts $\Delta \Phi_{1,2}$,
\beq
r_{1,2}=T^{-1}\ln|\lambda_{1,2}|; \ \ \Delta \Phi_{1,2} = -(\mathrm{arg} \lambda_{1,2} \mp \phi/2) .  
\label{RateEigen} 
\eeq

Equations (\ref{MainEqFourierc}) have a symmetry with respect to reflection of time: this pair of equations does not change after the following transformation: 
\beqa 
&& \tau \rightarrow \tau_0 - \tau;    \nonumber \\ 
&& c^+ \rightarrow c^- ; \ \ \  c^- \rightarrow -c^+ .   \nonumber
\eeqa
This $CT$-symmetry entails that the eigenvalues $\lambda_{1,2}$ are mutually inverse and that the eigenvectors $\mathbf{v}_{1,2}$ are orthogonal: 
\beq
\lambda_1 \lambda_2 = 1; \ \ \ \mathbf{v}_1 \cdot \mathbf{v_2}=v_{1}^{+} v_{2}^{+} + v_{1}^{-} v_{2}^{-}=0 .  \label{EigenVV} 
\eeq
This circumstance does not necessarily mean that only one of the two slipping-beam modes is unstable, since on top of these eigenvalues the same-beam factor $e^{i\mathsf{S}}$ contributes to the growth rate as well. However, the mode with $|\lambda|>1$ is more unstable, so it is reasonable to limit our attention to this mode only.   

Equations of motion~(\ref{MainEqFourierc}) select two special mode phases $\psi$, where the growth rate may be maximal. The first one is selected by the wake; it is the phase where the wake provides maximal interaction, i.e. where its Fourier images $\mathsf{S}$ and $\mathsf{O}$ reach their maxima. For example, a resonator wake with the frequency $\omega_r$ selects the resonating phase $\psi=-\omega_r \tau_0$; a thick-wall resistive wake selects the phase $\psi = -0$, where its images $\mathsf{S}$ and $\mathsf{O}$ go to infinity, with negative signs of their imaginary parts, etc. The second special phase, selected by Eqs.~(\ref{MainEqFourierc}), corresponds to a resonance between the beams, when the relative phase advance per the slippage time is a multiple of $2\pi$, i.e. 
\beq
\psi+\Delta \psi \equiv \phi + \chi = 2\pi n\,; \;\;\; n=0,\,\pm 1,\; \pm 2 ...  \,.
\label{ResBeams} 
\eeq
Whatever the chromatic phase $\chi$, there is one and only one beam-beam mode, corresponding to the resonance (\ref{ResBeams}), where the beam-beam interaction is enhanced. The phase parameter of this resonating mode may also be expressed as 
\beq
\phi_\mathrm{res} = -\chi \; (\!\!\!\!\! \mod 2\pi)\,.
\label{ResBeams2} 
\eeq
From here, one may generally conclude that it is beneficial to set the chromatic phase $\chi$, Eq.~\ref{chibb}, so that  
\beq
(\chi - \omega_b \tau_0) \; (\!\!\!\!\! \mod 2\pi)\, < 0. 
\label{ResBeams3} 
\eeq
In this case, the self-interaction is stabilizing at the beam-beam resonance, $\Im \mathsf{S} >0$, so the detrimental effect of the beam-beam resonance is reduced. Examples of that will be shown below.

\section{Resistive Wall}
  
\begin{figure}[h!]
\includegraphics[width=0.95\linewidth]{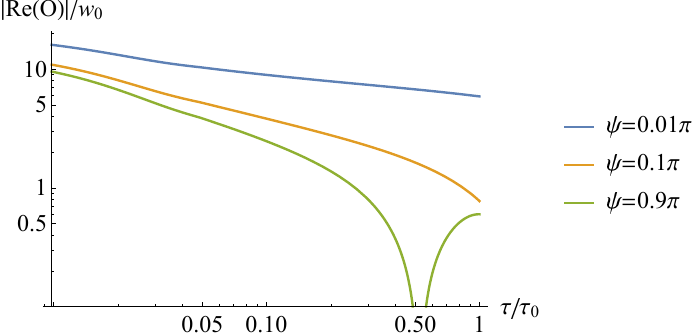}
\caption{\label{fig:ReO}
The real part of the function $\mathsf{O}(\psi, \tau)=w_0 \Upsilon (\psi,\tilde{\tau})$ is an even function of the mode parameter $\psi$.}
\end{figure}

\begin{figure}[h!]
\includegraphics[width=0.95\linewidth]{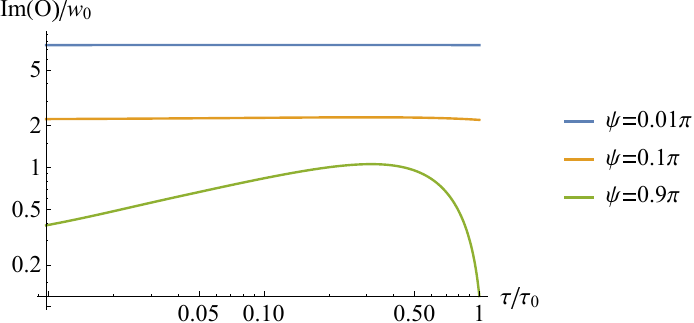}
\caption{\label{fig:ImO}
The imaginary part of the function $\mathsf{O}(\psi, \tau)$ is an odd function of the mode parameter $\psi$. The unstable modes are at $-\pi<\psi<0$.}
\end{figure}

\begin{figure}[h!]
\includegraphics[width=0.95\linewidth]{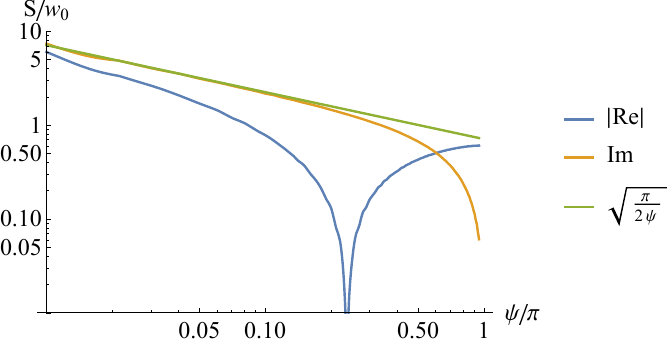}
\caption{\label{fig:Spsi}
Real (even) and imaginary (odd) parts of the same-beam growth factor $\mathsf{S}=\mathsf{O}(\tau_0)$ with their common low-phase asymptote. The unstable modes with $\Im{S}<0$ are at $-\pi<\psi<0$. Note that the imaginary part approaches its asymptote much sooner than the real one.}
\end{figure}

In this section the described method is applied to the case of a thick resistive wall, $W(s) \propto 1/\sqrt{s}$. Real and imaginary parts of the function 
\beqa
&& \mathsf{O}(\tau)= w_0 \sum_{n=0}^{\infty}\frac{\exp(i \psi(n + \tilde{\tau})}{\sqrt{n + \tilde{\tau}}} \equiv w_0 \Upsilon (\psi, \tilde{\tau}); \label{Otau}  \\
&&  \tilde{\tau} \equiv \tau/\tau_0 \nonumber 
\eeqa
are presented in Figs. \ref{fig:ReO} and \ref{fig:ImO}; the same-beam growth factor $\mathsf{S}$ is shown in Fig.\ref{fig:Spsi}.  At $|\psi| \ll 1$ the following approximations, found by the author, can be useful:
\beqa 
&& \mathsf{S}/w_0 = \Upsilon (\psi,1) \approx \sqrt{\frac{\pi}{2|\psi|}}(1+ i \mathrm{sgn}\psi)-1.45 - 0.66i \frac{\psi}{\pi} ; \nonumber \\
&& \Upsilon (\psi,\tilde{\tau}) \approx \Upsilon (\psi,1)+\frac{1}{\sqrt{\tilde{\tau}}}-\tilde{\tau}-0.5|\psi| \tilde{\tau} (1-\tilde{\tau}).   \label{Oasympt} 
\eeqa
This approximation for the function $\mathsf{S}(\psi)$ is especially remarkable: for all $|\psi| \leq \pi/2$ it is valid within the accuracy of 1\% or better. For the function $\mathsf{O}(\psi,\tau)$, the same accuracy is reached only at $|\psi| \leq 0.1$.

For long coupled-bunch waves, $|\psi| \ll 1$, the dependence of the interaction function $\Upsilon(\psi,\tilde{\tau})$ on the slipping phase $\tilde{\tau}$ can be neglected, so the resulting coupled-bunch modes turn out to be identical to the conventional coupled-bunch modes of the doubled beam with $2M$ bunches. However, for the short waves, $|\psi| \geq 1$, such reduction does not take place. Without slippage, the phase $\tilde{\tau}$ is frozen, while the coupled-bunch interaction depends on its value. Thus, collective tune shifts for the short waves depend on the specific frozen value of $\tilde{\tau}$. When the beams are slipping, this phase is running, and the collective tune shifts result from the proper integration over that. From here, one may conclude that the whole approach of this paper assumes that the growth rates of the short waves do not exceed the slippage period $T$ by much.  

For a round vacuum chamber with the circumference $C_0$, aperture radius $d$, conductivity $\sigma$, the conventionally normalized transverse wake function is \cite{chao1993physics}:
$$
W_\perp(\tau)=W_0 \sqrt{\frac{\tau_0}{\tau}}; \ \ \ W_0=\frac{2}{\pi} \frac{C_0}{d^3} \frac{1}{\sqrt{\sigma \tau_0}}.
$$   
For a flat chamber, the thick-wall resistive wake is reduced by the Yokoya factors $\pi^2/12$ vertically and $\pi^2/24$ horizontally. The dimensionless slipping-beam intensity parameter $w_0$ can be written,
$$
w_0=\frac{N_b r_0 W_0 \beta}{2 \gamma} \frac{T}{T_0}=\frac{N_b r_0 W_0 \beta}{2 \gamma M |\eta \delta p/p|}. 
$$  
Here $N_b$ is the bunch population, $r_0$ is the classical radius, $\beta$ is the average beta-function, and $\gamma$ is the relativistic factor. 

In general, the spectrum of slipping-beam modes is determined by two dimensionless values: the intensity parameter $w_0$, and the beam-beam phase shift $\Delta \psi =\chi - \omega_b \tau_0$. If the latter is small enough, $|\Delta \psi| \ll 1$, the results are almost the same as for its zero value, so only one parameter, $w_0$, remains. 

In the next Section, the example of the Fermilab Recycler Ring (RR) is considered, first, for zero beam-beam phase shift, $\Delta \psi=0$, and then it will be shown how chromaticity may change the results.   

\section{Slipping-Beam Modes at the Recycler Ring} 

For the RR with $C_0=3.3$km, $\gamma=9.5$, $\beta=22$m, $\eta=-0.0087$, $\delta p/p=0.0027$, and with Proton Improvement Plan II values $N_b=7.6 \cdot 10^{10}$ and $M=504$, the slipping-beam intensity parameter $w_0$ comes out as $w_0=0.12$. For that number of bunches and the betatron tune $Q_y=24.4$, the bunch-to-bunch phase advance $\phi_{bb} = 0.30$ and bunch-to-bunch slipping time $T/T_0=(M |\eta \delta p/p|)^{-1} =90$ revolutions.

\subsection{Zero beam-beam phase shift, $\Delta \psi=0$}

\begin{figure}[h!]
\includegraphics[width=0.95\linewidth]{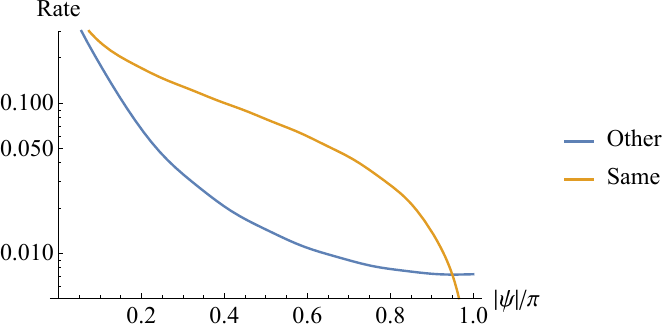}
\caption{
\label{fig:RateLog}
Growth rate in units of $1/T$, i.e. $r_1 T$ of Eq.~(\ref{RateEigen}), for the specified parameters of the RR versus the coupled-bunch phase $-\pi \leq \psi < 0$. The yellow line gives the contribution of the same beam, $\Im \mathsf{S}$, while the blue one shows the growth rate caused by the other beam. The total rate (not shown in this figure) is the sum of the two. The beam-beam phase shift is zero, $\Delta \psi=0$
}
\end{figure}

The growth rate versus the negated coupled-bunch phase is shown in Fig.~\ref{fig:RateLog}. 
It is clear that the {\it same} beam interaction $\mathsf{S}$ dominates the {\it other} one $\mathsf{O}$  when the mode phase $\psi$ is sufficiently large. The condition for this dominance can be estimated from Eqs.~(\ref{CoastFreq}, \ref{MainEqFourierc}) 
$$
|\psi|\gg w_0^{2/3} (2\pi)^{1/3}\,,
$$
yielding $|\psi|/\pi \gg 0.14$ for the RR parameters above, which agrees with Fig.~\ref{fig:RateLog}.  

\begin{figure}[h!]
\includegraphics[width=0.95\linewidth]{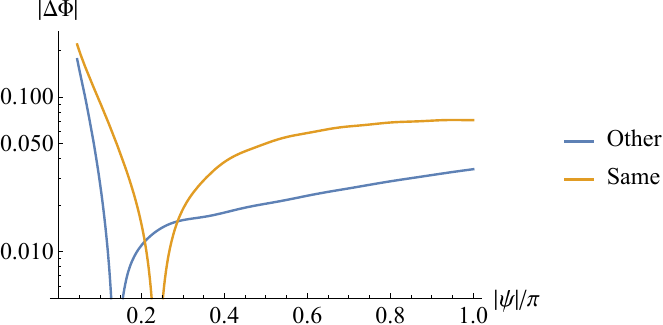}
\caption{
\label{fig:dPhaseAdvanceLog}
Phase advance shift $\Delta \Phi$ of the unstable mode. It is negative at small phases, and then it changes the sign. The beam-beam phase shift is zero, $\Delta \psi=0$.
}
\end{figure}

\begin{figure}[h!]
\includegraphics[width=0.95\linewidth]{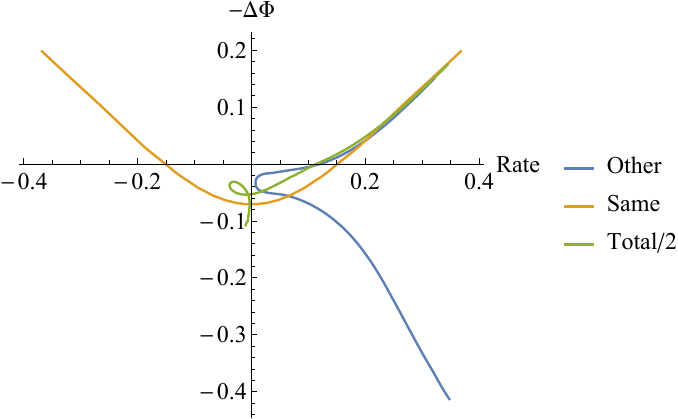}
\caption{
\label{fig:RatePhaseParPlot3}
Parametric plot of the total phase advance shift versus total growth rate, as well as their two contributions, {\it{same}} and {\it other}. The phase $\psi$ changes from $-0.05\pi$ (upper right corner) to $-1.95\pi$. For a more convenient comparison, the {\it total} values are divided by 2.    
}
\end{figure}

Generally, the conventional coupled-bunch growth rate is exactly zero for $|\psi|=\pi$ and any sort of wake. As one can see in Fig.~\ref{fig:RateLog}, this is not the case for the slipping beams: although at $|\psi|=\pi$ the growth rate is low compared with its values at small coupled-bunch phases, it is still not zero. For $w_0 \ll 1$, this rate is well fitted by $r \approx 0.5 w_0^2$.
Figure \ref{fig:dPhaseAdvanceLog} demonstrates the phase advance shift of the unstable mode, which may be important for Landau damping if the phase advance shift exceeds the growth rate, or is at least comparable to it. The parametric plot presented in Fig.\ref{fig:RatePhaseParPlot3} can be used to determine whether or not that is the case. While for small phases $|\psi| \leq 1$ the entire phase advance shift $|\Delta \Phi+\Re \mathsf{S}|$ is comparable with the total growth rate $r-\Im \mathsf{S}$, closer to $|\psi|=\pi$ the phase advance shift may be much higher than the growth rate.    

\begin{figure}[h!]
\includegraphics[width=0.95\linewidth]{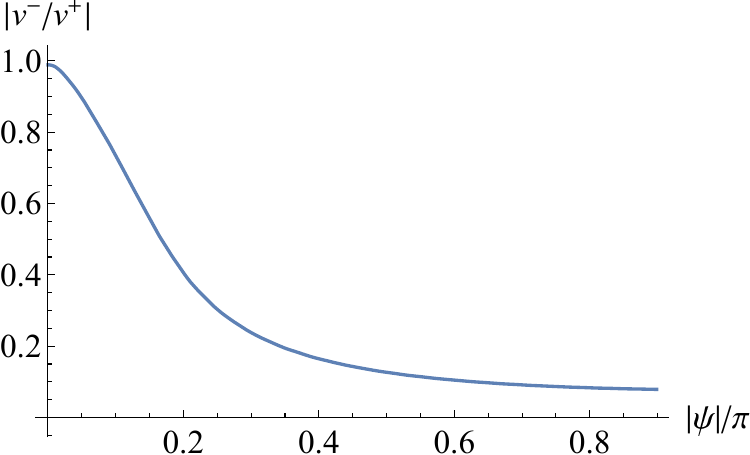}
\caption{
\label{fig:VNPPlot}
Modulus of the ratio of the eigenvector components. 
}
\end{figure}

\begin{figure}[h!]
\includegraphics[width=0.95\linewidth]{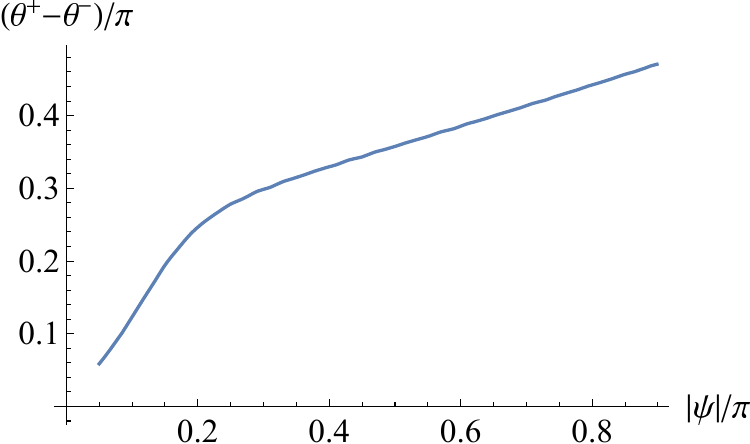}
\caption{
\label{fig:APNPlot}
Relative argument of the eigenvector components in the units of $\pi$, $(\mathrm{arg}(v^+) - \mathrm{arg}(v^-))/\pi$. The beam-beam phase shift is zero, $\Delta \psi=0$, as for all the figures above. 
}
\end{figure}

Two plots for the eigenvectors are given in Figs. \ref{fig:VNPPlot} and \ref{fig:APNPlot}.

\subsection{The coupled-beam spectrum for arbitrary chromaticity}

\begin{figure}[h!]
\includegraphics[width=0.95\linewidth]{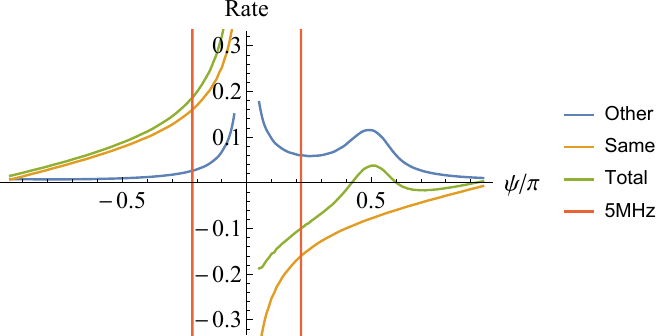}
\caption{
Growth rates for the coupled-beam modes versus their phase number $\psi$ for the RR parameters. The beam-beam phase shift $\Delta \psi = -0.5 \pi$, corresponding to the chromaticity $\xi = -0.8$. Note the beam-beam resonance at $\psi=0.5 \pi$, in agreement with Eq.~(\ref{ResBeamsPsi}). Red vertical lines show relation between the phase $\psi$ and the frequency $f=\psi/(2\pi \tau_0)$.   
}
\label{Fig:RateChiMinus0p5pi}
\end{figure}

\begin{figure}[h!]
\includegraphics[width=0.95\linewidth]{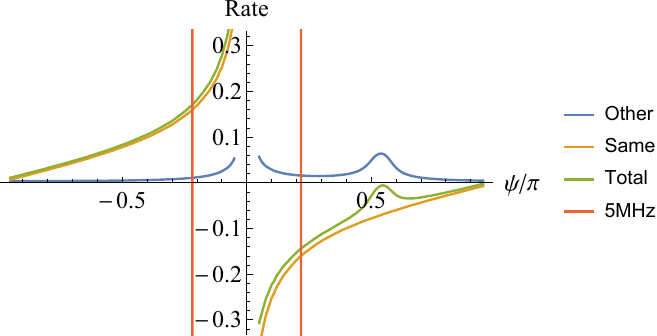}
\caption{
The same as Fig.~\ref{Fig:RateChiMinus0p5pi}, but with the beam-beam phase shift $\Delta \psi =1.5 \pi$, i.e. $2\pi$ larger than there. The resonance location is same, but its strength dropped.  
}
\label{Fig:RateChiPlus1p5pi}
\end{figure}

\begin{figure}[h!]
\includegraphics[width=0.95\linewidth]{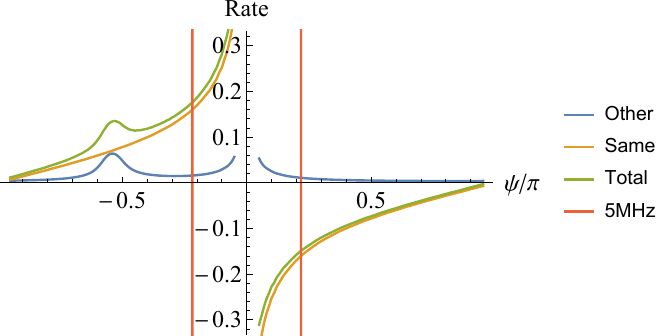}
\caption{
Growth rates for $\Delta \psi = -1.5 \pi$, leading to the mirror-symmetric beam-beam contribution in comparison with Fig.~\ref{Fig:RateChiPlus1p5pi}. The corresponding RR chromaticity $\xi=-2.8$. 
}
\label{Fig:RateChiMinus1p5pi}
\end{figure}

At zero beam-beam phase shift, $\Delta \psi \equiv \chi - \omega_b \tau_0=0$, the beams are in resonance for the same mode phase $\psi$, where the interaction functions $\mathsf{S}$ and $\mathsf{O}$ are maximal, i.e. at $\psi=-0$. However, this would not be so for arbitrary chromaticity, as it was discussed at the end of Section~\ref{Sec:MainEq}. For a given chromatic factor $\chi$, Eq.~(\ref{chibb}), the resonant coupled-beam mode phase $\psi$ is found to be 
\beq
\psi_\mathrm{res}=-\Delta \psi \; (\!\!\!\!\! \mod 2\pi)=(-\chi + \omega_b \tau_0) \; (\!\!\!\!\! \mod 2\pi)\,,
\label{ResBeamsPsi} 
\eeq
assuming, as everywhere above, $|\psi| \leq \pi$. While the inter-bunch betatron phase $\omega_b \tau_0$ is given by the beam structure, the chromaticity is a variable parameter, normally used to make the beams more stable. Changing the chromatic phase $\chi$ moves up and down the resonant mode $\psi_\mathrm{res}$. Its optimal position depends on the damper bandwidth and should be considered with the intra-bunch head-tail motion taken into account; in its completeness, the latter problem lies outside the framework of this article. To illustrate how the coupled-beam growth rates can be influenced by chromaticity, Eqs.~(\ref{MainEqFourierc}) have been solved for the RR parameters and different chromaticities; the growth rates in units of the inverse slipping period $1/T$ are presented in Figs.~\ref{Fig:RateChiMinus0p5pi}~--~\ref{Fig:RateChiMinus1p5pi}.  The first of them, Fig.~\ref{Fig:RateChiMinus0p5pi}, shows that while the beam-beam phase shift generally suppresses beams interaction, it makes the resonance mode $\psi \approx 0.5\pi = -\Delta \psi$ unstable. Figure~\ref{Fig:RateChiPlus1p5pi} demonstrates that the positive sign of chromaticity below transition is not necessarily worse than its negative sign, from the beam-beam instability point of view. The same conclusion is additionally supported by Fig.~\ref{Fig:RateChiMinus1p5pi}.  While a difference of the chromaticity $\xi$ by a couple of units considerably changes the beam-beam interaction in the RR, as one may conclude from comparison of Fig.~\ref{Fig:RateChiMinus0p5pi} and Fig.~\ref{Fig:RateChiMinus1p5pi}, this difference corresponds to just a tiny value of the single-bunch head-tail phase $\zeta = \xi \sigma_p/ Q_s$, where $\sigma_p$ is the relative rms momentum spread within the bunch, and $Q_s=\omega_s/\omega_0$ is the synchrotron tune.   

\section{\label{sec:Distribution} Distribution of Coupled-Bunch Rates over the Head-Tail Modes}

Let us imagine now that the coupled-bunch and beam-beam coherent tune shifts are all found, and ask how are they are distributed between the head-tail modes. In those cases when the wake fields of preceding bunches can be considered constant at the bunch length, i.e. {\it flat}, the problem is reduced to a single-bunch dynamic equation where the coupled-bunch forces are taken into account in the same way as the bunch-by-bunch damper, see e.g. \cite{burov2009head, PhysRevSTAB.15.054403, Burov:2013nb, burov2014nested}. Below, the problem is considered for two limit cases: for zero space charge (ZSC), $\Delta Q_\mathrm{sc} \ll Q_s$, and for the strong space charge (SSC), $\Delta Q_\mathrm{sc} \gg Q_s$, with $\Delta Q_\mathrm{sc}=\Delta \omega_\mathrm{sc}/\omega_0$ as the space charge tune shift at the bunch center. In both cases, only coupled-bunch wake forces will be taken into account, and they will be assumed flat on the bunch length, while the intra-bunch wakes will be neglected. 

\begin{figure}[htb]
\includegraphics[width=0.95\linewidth]{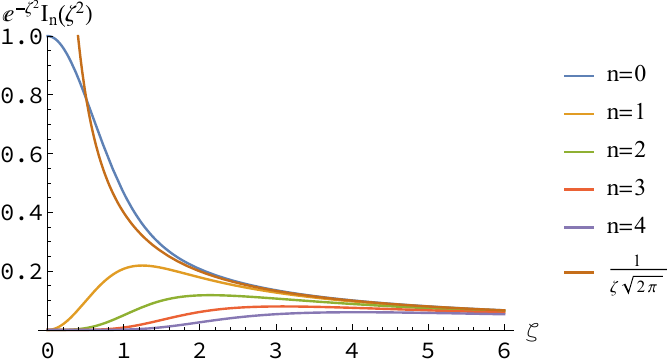}
\caption{
\label{fig:NHTFormFactor}
ZSC (zero space charge) dipole formfactors $F_n=e^{-\zeta^2} \mathrm{I}_n(\zeta^2)$ for a Gaussian bunch versus the rms head-tail phase $\zeta$; $\mathrm{I}_n$ is the modified Bessel function.  
}
\end{figure}

For the ZSC case, following Ref.~\cite{burov2014nested}, the intra-bunch pattern of a mode can be expanded over the {\it{nested head-tail}} basis in the synchrotron phase space
$$ 
\Psi_{l\alpha} = \exp(il\varphi + i\zeta_\alpha \cos \varphi),
$$
where $l=0,\pm 1, \pm 2,...$, $\varphi$ is the synchrotron phase, $\zeta_\alpha$ is the chromatic head-tail phase at the radial position $\alpha$. Components of that expansion $X_{l\alpha}$ satisfy the following equation: 
\beqa
&& (\nu -l \omega_s) X_{l\alpha} = \Delta \omega  i^{-l} \mathrm{J}_l(\zeta_\alpha) \bar{X}; \label{NHTEq}\\
&& \bar{X}= n_r^{-1}\sum_{m \beta} i^{m}\mathrm{J}_m(\zeta_\beta) X_{m\beta}, \nonumber
\eeqa
where $\Delta \omega = \Delta \Phi + i r$ is the complex frequency shift of the coupled-bunch wake and point-like bunches, $n_r$ is the number of the radial rings representing the bunch in the longitudinal phase space (ideally $n_r \rightarrow \infty$); 
$$
\mathrm{J}_l(\zeta_\alpha)=\frac{ i^{-l}}{2\pi}\int_0^{2\pi} \exp(il\varphi + i\zeta_\alpha \cos \varphi) d\varphi
$$ 
is the Bessel function as the dipole moment of the basis function $\Psi_{l\alpha} $ and $\nu$ is the sought-for eigenvalue. Note that for a rigid-bunch motion, when the bunch moves as a whole with the amplitude $\bar{X}$, 
$$
X_{l\alpha} = i^{-l} \mathrm{J}_l(\zeta_\alpha) \bar{X}.
$$
From Eq.(\ref{NHTEq}), a dispersion equation on the eigenvalues $\nu$ follows: 
\beqa
&&  \Delta \omega \sum_{l}\frac{F_l}{\nu-l \omega_s} = 1; \label{DispEqGen} \\
&& F_l =  \int_0^{\infty} \mathrm{J}_l^2(\zeta r) f(r) rdr; \ \ \int_0^{\infty}  f(r) rdr= 1.  
\eeqa
where $f(r)$ is the normalized longitudinal phase space density, and the values $F_l(\zeta)$ will be called the head-tail or dipole formfactors. Note that 
$$  
\sum_{l=-\infty}^{\infty} F_l(\zeta) =1
$$
for any chromatic factor $\zeta$ and any distribution function $f(r)$. For the Gaussian distribution, $f(r)=e^{-r^2/2}$, the formfactor integrals can be analytically taken:
\beq
F_l(\zeta)= e^{-\zeta^2} \mathrm{I}_l(\zeta^2), \label{NHTFFactor}
\eeq
where $\mathrm{I}_l$ is the modified Bessel function. Some of these formfactors are shown in Fig.\ref{fig:NHTFormFactor}.
Roughly speaking, if the head-tail phase $|\zeta| > 1$, the ZSC formfactors $F_l$ up to $|l| \simeq |\zeta|$ are close to their common asymptotic value $(2\pi \zeta^2)^{-1/2}$, while those of the higher modes could be neglected.  

The eigenvalues $\nu$ are easily found in two opposite limit cases, for small and large values of the coupled-bunch frequency shift $\Delta \omega$. If it is small compared with the synchrotron frequency, $|\Delta \omega| \ll \omega_s$, all the unperturbed collective frequencies $l\omega_s$ just slightly shift, sharing the total coherent shift according to their form-factors $F_l$:
\beq
\nu_l = l\omega_s + \Delta \omega F_l(\zeta) . \label{NHTEValsPert}
\eeq
In the opposite limit, when $|\Delta \omega| \gg \omega_s |k -m|$ for all $k,m$ with non-negligible form-factors, i.e. for $ |\Delta \omega| \gg \omega_s |\zeta|$, Eq.(\ref{DispEqGen}) shows that a single eigenvector essentially takes the entire tune shift $\Delta \omega$. This dominant mode is nothing but the rigid-bunch motion, while all other eigenvectors are of very small dipole moment. Thus, the high coupled-bunch tune shift gives rise to the rigid-bunch motion; note that notwithstanding the chromaticity, there is no chromatic traveling wave in that powerful mode. As to the other eigenvectors, for such a high coupled-bunch tune shift, each of them becomes of almost zero dipole moment. Thus, growth of $|\Delta \omega|$ from very small to very high values first leads to the proportional growth of all the head-tail tune shifts, but as soon as the $|\Delta \omega|$ becomes comparable with the band of the participating harmonics, the common rigid-bunch mode is formed, tending to take the entire tune shift $\Delta \omega$. 

Although that lies outside the scope of this paper, it is still worth noting that in the limit of a very large number of terms in Eq.(\ref{DispEqGen}), its sum can be replaced with an integral, and the equation transforms into a conventional dispersion equation of a medium, consisting of many harmonic oscillators affected by a collective force proportional to their common dipole moment. In that case, the transfer from the perturbed intrabunch modes to the powerful rigid-bunch mode, which happens at $|\Delta \omega| \simeq |\zeta| \omega_s$, is similar to the appearance of a discrete common mode above a continuous incoherent van Kampen spectrum and the loss of Landau damping \cite{van1957dispersion, Burov:2011zza}.     

Now let us see how the coupled-bunch modes are shared between the intra-bunch modes in the case of the strong space charge, SSC. For the RR, the space charge is strong: the maximal space charge frequency shift $\Delta \omega_\mathrm{sc}$ exceeds both the synchrotron frequency $\omega_s$ and the coherent frequency shifts $\Delta \omega$ by about an order of magnitude. For the strong space charge, the head-tail degree of freedom becomes one-dimensional; its eigensystem can be found from the ordinary integro-differential equation of Ref.\cite{burov2009head}. Being expanded over the orthonormal basis of the space charge harmonics $y_k^0$, $k=0,1,2...$, this equation is reduced into the standard algebraic eigensystem problem. When the wake fields are dominated by flat coupled-bunch (and possibly feedback) terms, this set of linear homogeneous equations on the eigenfunction components $B_k$ can be written,
\beqa 
&& (\nu-\nu^0_k) B_k = \Delta \omega I_k \sum_m B_m I_m^{*}; \label{BurovSCE} \\
&& I_k=\int_{-\infty}^{\infty} e^{i \zeta s} \rho(s) y_k^0(s) ds, \label{SCIk}
\eeqa
where $\rho(s)$ is the bunch normalized line density, $\int_{-\infty}^{\infty} \rho(s) ds = 1$,  $\nu$ is the eigenvalue to be found and $\nu^0_k \simeq k^2 \omega_s^2/\Delta \omega_\mathrm{sc}$ is its $k$-th no-wake value. The {\it{italicized}} symbol $I_k$ for the dipole moments of the basis functions is not to be confused with the straight one used for the modified Bessel functions $\mathrm{I}_k$.

Eq.(\ref{BurovSCE}) can be solved similarly to the zero space charge case; a dispersion equation for the sought-for eigenvalues $\nu$ follows:
\beq
\Delta \omega \sum_m \frac{|I_m|^2}{\nu-\nu^0_m} = 1 . \label{EValsEq}
\eeq 

\begin{figure}[htb]
\includegraphics[width=0.95\linewidth]{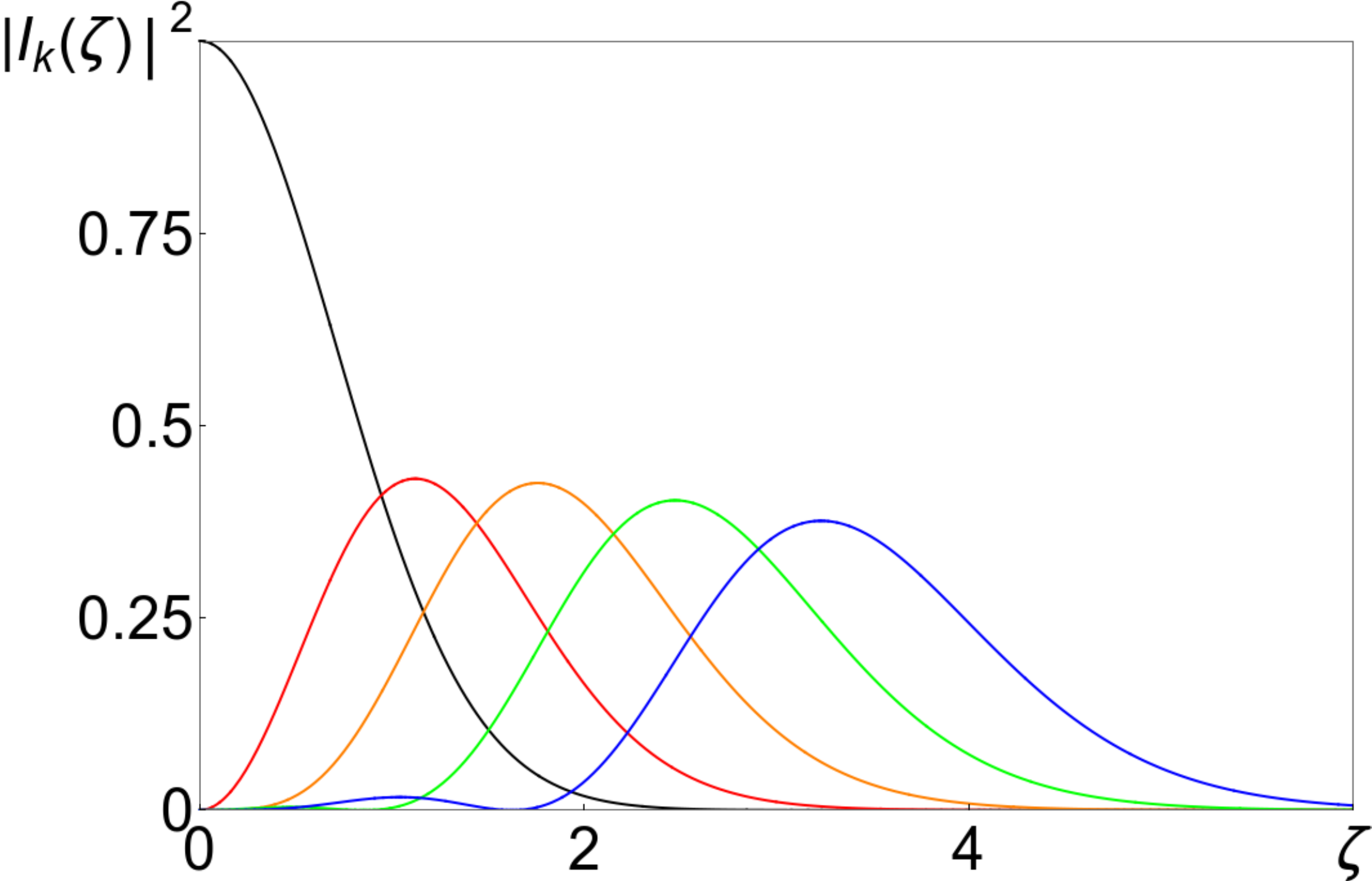}
\caption{
\label{fig:Ik2_vs_zeta}
Chromatic formfactors of the coupled-bunch contributions to head-tail modes $k=0,1,2,3,4$ (black, red, orange, green and blue) at strong space charge for a Gaussian bunch.  
}
\end{figure}

Formally, this equation is of the same type as its counter-part for the ZSC case, Eq.(\ref{DispEqGen}). The dipole formfactors now are $F_l=|I_l|^2$ since the dipole moments of the basis functions are $I_l(\zeta)$ for SSC, instead of $i^l \mathrm{J}_l(\zeta_\alpha)$ for ZSC. Thanks to orthonormalization of the basis, 
\beq 
 \int_{-\infty}^{\infty} y_k(s) y_l(s) \rho(s) ds = \delta_{kl}, \label{ykorthonorm} \\ 
\eeq
it is true that
\beq
\sum_k |I_k|^2=1. \label{sumIk2}
\eeq
For a Gaussian bunch, these functions are presented in Fig.\ref{fig:Ik2_vs_zeta}. Similarly to the ZSC case, SSC formfactor of the $k$-th mode reaches its maximum at the head-tail phase $|\zeta| \simeq k$, being insignificant even a few units below that value. However, the SSC formfactor behaves differently above its maximum. While for ZCS all the non-negligible formfactors follow the same asymptotic $\propto |\zeta|^{-1}$, the SSC ones exponentially decay soon after reaching their maxima. So, for any chromaticity there are not more than 2 to 4 SSC harmonics, which are sufficient to be taken into account. One more important difference is that for SSC the distance between the neighbor unperturbed lines, $\nu_k^0$ and $\nu_{k+1}^0$, grows $\propto k$, while for ZSC this distance is constant. As a result, the threshold of the rigid-bunch mode in both cases is proportional to chromaticity. While for the ZSC case this threshold is $|\Delta \omega| \simeq |\zeta| \omega_s$, for the SSC one it is $|\Delta \omega| \simeq |\zeta| \omega_s^2/\Delta \omega_{\mathrm{sc}} \ll |\zeta| \omega_s$.     

Another stabilizing effect of the chromaticity relates to Landau damping. The rigid-bunch mode is known to not have any Landau damping. If this mode is not formed, i.e. if the coupled-bunch tune shift does not exceed the distance between neighboring head-tail modes, it excites them independently according to their formfactors. Thus, every participating head-tail mode is Landau-damped with its no-wake rate~\cite{burov2009head, PhysRevSTAB.18.074401}, assuming the single-bunch wake to be small enough. A consequence of that is very strong dependence of the intrinsic Landau damping on the chromaticity for the SSC case, as fast as $\propto \zeta^4$, so a sufficiently high chromaticity should suppress the instability. For ZSC, the higher harmonics contribute more to Landau damping from the longitudinal degree of freedom.

For both the ZSC and SSC cases, i.e for both Eq.(\ref{NHTEq}) and Eq.(\ref{BurovSCE}), the following general theorems can be proven (see the Appendix):
\begin{enumerate}
\item The sum of the head-tail tune shifts driven by the coupled-bunch interaction is equal to the total coupled-bunch tune shift, $\sum_k (\nu_k - \nu_k^0) = \Delta \omega .$ For the ZSC case, $\nu_k^0 = k \omega_s$, and $k=0, \pm1, \pm2,...$, while for SSC $\nu_k^0 \simeq k^2 \omega_s^2/\Delta \omega_\mathrm{sc}$, and $k=0, 1, 2,...$. 

\item The sign of each head-tail growth rate caused by the coupled-bunch interaction is the same as the sign of the total coupled-bunch growth rate, $0 \leq \Im \nu_k /\Im \Delta \omega \leq 1 .$  This theorem limits maximal growth rate of each head-tail mode and denies the possibility for a resistive damper to cause an instability of any of them; as soon as $\Im \Delta \omega \leq 0$, for all partial growth rates $\Im \nu_k \leq 0$ as well. 

%This conclusion is generally not valid when single bunch wakes are taken into account, see Refs.~\cite{burov2014nested, PhysRevAccelBeams.19.084402, Burov:2018pjl}. 

\item For a purely reactive damper, when $\Im \Delta \omega =0$, the head-tail eigenvectors are real and orthogonal, $\mathbf{B}^p \cdot \mathbf{B}^q =\delta_{pq}$, and the same goes for $\mathbf{X}$. Otherwise they are, generally, neither real nor orthogonal. This theorem may be useful for the analysis of perturbations of the head-tail modes formed by a strong reactive damper.

\end{enumerate}

Strictly speaking, these theorems become invalid as soon as single-bunch wakes are taken into account, but various problems of such sort are beyond the scope of this article; some examples of common action of a damper and a single-bunch wake are presented in Refs.\cite{burov2014nested, PhysRevAccelBeams.19.084402, Burov:2018pjl}. However, if the single-bunch wake is sufficiently small, it can be taken into account as a perturbation of the coupled-bunch eigensystem $\{ \mathbf{B}\}$. To do that, the Hermitian adjoint set of the eigenvectors $\tilde{\mathbf{B}}$ corresponding to complex-conjugate tune shift $\Delta \omega^*$ has to be computed. As soon as it is done, the diagonal matrix elements of the single-bunch wake  $\langle{p}| \mathcal{W} |{p} \rangle \equiv \tilde{\mathbf{B}}^{(p)} \cdot \mathcal{W} \mathbf{B}^{(p)} $ provide the tune shifts through the presumably small perturbation represented by the single-bunch matrix 
\beq
\mathcal{W}_{mn}= \int \int W(s-s') e^{i \zeta (s-s')} \rho(s) \rho(s') y_m^0(s) y_n^0(s') ds ds'. \label{SBWake}
\eeq
The assumption that the single-bunch wake is small is justified if the absolute value of its tune shift $\propto |\langle{p}| \mathcal{W} |{p} \rangle |$ is much smaller than the distance between the neighbor tunes $|\nu^{(p)}-\nu^{(p \pm 1)}|$.  

The coupled-bunch instability can be considered independently of the single-bunch wakes as soon as the coupled-bunch wake forces exceed the single-bunch ones. For the resistive wall wake, this condition is satisfied if the coupled-bunch mode phase number $\psi$ is sufficiently small. To quantify this condition, the coupled-bunch kick $\sim \mathsf{S}$ can be compared with the kick that the bunch in the rigid mode gets from its own wake. Using the low-frequency approximation, $|\psi| \ll 1$, for the coupled-bunch factor $\mathsf{S}$ and taking the average of the single-bunch wake along the Gaussian bunch in the rigid-bunch mode, one gets the condition for the coupled-bunch dominance:  
\beq
\frac{\mathsf{S}(\psi)}{\sqrt{\tau_0}} \approx \sqrt{\frac{\pi}{2 |\psi| \tau_0}} > \frac{1}{\sqrt{2\sigma_\tau}}, 
\eeq
where $\sigma_\tau$ is the rms bunch length. From here, the upper limit on the mode phase $\psi$ follows,
\beq
|\psi| < \psi_\mathrm{SB} \equiv \pi \sigma_\tau/\tau_0 \,.
\label{psiSB}
\eeq 
This condition can be also expressed in terms of the coupled-bunch frequency $f=|\psi|/(2\pi \tau_0)$: 
\beq
f < f_\mathrm{SB} \equiv M f_0 \sigma_\tau/(2\tau_0) \,,
\label{fSB}
\eeq
with $f_0$ as the revolution frequency. For the RR $M=500$ bunches, which rms length $\sigma_\tau=40$~cm, the last condition results in $f < 1.6$~MHz. From here one may conclude that a feedback damper with the bandwidth $\simeq 2$~MHz would effectively suppress the related band of the coupled-bunch modes, while effectiveness of further broadening of the damper bandwidth requires a special consideration, where the single-bunch wakes cannot be neglected. As it was shown in Ref.~\cite{Burov:2018pjl}, the combined action of the damper and a single-bunch wake may lead to a special absolute-convective instability, if the single-bunch wake amplitude is comparable to its ZSC TMCI threshold value. The question of how far this limit is for the current and planned intensity of the RR lies outside the scope of this paper.  

Since that sort of low-frequency damper was proposed in the first version of this paper~\cite{Burov:2016jsh}, the specified low-frequency damper was designed and installed in the RR~\cite{Eddy:2018tdm}, which allowed the chromaticity to be reduced from -20 down to -7. As a result, the total loss was reduced by a factor of almost two with losses at the Abort and Muon Extraction Lambertsons reduced significantly. The ability to run with much lower chromaticity also provided much more freedom in choosing the working point and to remove the injection phase offsets~\cite{Ainsworth:2018HB}.

\section{Summary}

Coupled-bunch modes are described for two slipping beams in a storage ring, as it is the case in the Fermilab Recycler Ring. 
Distribution of the inter-bunch growth rates between the intra-bunch head-tail modes is considered. Possibilities to stabilize the beams by means of chromaticity and feedback are shown. 

Since this paper was written (in June 2016), the proposed low-frequency damper has been built, installed~\cite{Eddy:2018tdm} and has demonstrated its effectiveness~\cite{Ainsworth:2018HB}. 

I am thankful to S. Nagaitsev for his encouraging interest in this problem and to T. Zolkin for useful discussions and technical help.

FNAL is operated by Fermi Research Alliance, LLC under Contract No. DE-AC02-07CH11359 with the United States Department of Energy.

\appendix

\section{\label{secAP:Theorems} Theorems on distribution of the coupled-bunch tune shift over the intra-bunch modes}

Here the three theorems mentioned in Section \ref{sec:Distribution} are proved. Since the proofs are similar for ZSC and SSC, only the latter case is presented.   

\subsection{Theorem of the sum of tune shifts}

The dispersion equation (\ref{EValsEq}) can be transformed into a polynomial one by 
multiplying it with the product $\prod_k (\nu - \nu^0_k)$. The leading coefficient of the resulting polynomial equation is 1, and the next one is $-\Delta \omega - \sum_k \nu_k^0$. Due to a general property of polynomial equations, the negated value of the latter is the sum of the roots $\nu_k$. Thus, the sum of the head-tail tune shifts $\nu_k-\nu_k^0$ is equal to the total tune shift:
\beq
\sum_k (\nu_k-\nu_k^0) = \Delta \omega . \label{TuneShiftSum}
\eeq   
Since the unperturbed tunes are all real, it follows that sum of all the head-tail growth rates is the coupled-bunch growth rate 
\beq
\sum_k \Im \nu_k = \Im \Delta \omega . \label{RateSum}
\eeq

\subsection{Theorem of the growth rate signs} 

Let's prove that all the growth rates $\Im \nu_k$ have the same sign; in other words, the intra-bunch modes are either all stable or all unstable, depending on the sign of the total growth rate $\Im  \Delta \omega$. First, let's slightly rewrite Eq.(\ref{BurovSCE}):
\beq 
\nu B_k = \nu^0_k B_k + \Delta \omega I_k \sum_m B_m I_m^{*} . \label{nuBk}
\eeq
Then, this equation can be multiplied by $B_k^*$ and summed over all the components $k$, resulting in,
\beq 
\nu  = \bar{\nu^0}  + \Delta \omega |\bar{Y}|^2 , \label{HTTune}
\eeq
where 
$$
\bar{Y} \equiv \sum_k B_k I_k^* \equiv \mathbf{B} \cdot \mathbf{I}
$$
is the dipole moment associated with the normalized eigenvector $\mathbf{B}$, 
$$
\mathbf{B} \cdot \mathbf{B} \equiv  \sum_k B_k B_k^* = 1,
$$ 
and $\bar{\nu^0} \equiv \sum_k \nu^0_k |B_k|^2$. Taking the imaginary part of Eq.(\ref{HTTune}) leads to
\beq
\frac{\Im \nu}{\Im \Delta \omega} = |\bar{Y}|^2 \geq 0,
\eeq
which proves the theorem. Due to the Cauchy inequality, 
$$ 
|\bar{Y}|^2  \leq 1.
$$ 
By virtue of Eq.(\ref{RateSum}), the sum of the dipole formfactors $|\bar{Y}|^2$ over all the eigenvectors (distinguished by the superscript $^{(p)}$) is equal to 1:
\beq
\sum_p |\bar{Y}^{(p)}|^2 \equiv \sum_p |\mathbf{B}^{(p)} \cdot \mathbf{I}|^2 = 1.
\eeq
Thus, one single intra-bunch mode can take the entire tune shift $\Delta \omega$ if and only if its eigenvector is identical to the dipole moment vector, $\mathbf{B} =\mathbf{I}$, which means the mode has to be the rigid-bunch one. That happens when the value of the total tune shift exceeds the bandwidth of the harmonics involved. In that case, all other modes have a negligibly small dipole moment and share almost no growth or damping rate from the coupled-bunch interaction.

\subsection{Theorem of the orthogonality of eigenvectors}

This subsection formulates a sufficient condition for the eigenvectors' orthogonality. To do that, we have to deal simultaneously with two different eigenvectors; to distinguish them, the parenthesized superscripts will be used. Let us write Eq.(\ref{nuBk}) for an eigenvector $\mathbf{B}^{(p)}$ and multiply it, left to right, by the eigenvector  $\mathbf{B}^{(q)}$: 
$$
\nu^{(p)} \mathbf{B}^{(p)} \cdot \mathbf{B}^{(q)} = \sum_l \nu_l^0 {B}^{(p)}_l {B}^{(q)*}_l + \Delta \omega \bar{Y}^{(p)} \bar{Y}^{(q)*} ,  
$$
with $\bar{Y} \equiv \mathbf{B} \cdot \mathbf{I}$ as the dipole moment of the eigenvector $\mathbf{B}$.
The same equation (\ref{nuBk}) can be written for the eigenvector $\mathbf{B}^{(q)}$, which can then be multiplied, right to left, by the eigenvector $\mathbf{B}^{(p)}$. After that, the second equation can be subtracted from the first, yielding,
\beq
(\nu^{(p)} -\nu^{(q)*}) \mathbf{B}^{(p)} \cdot \mathbf{B}^{(q)} = 2 i \Im \Delta \omega \bar{Y}^{(p)} \bar{Y}^{(q)*}. \label{orthoB}
\eeq
Therefore, as soon as the coupled-bunch tune shift is real, $\Im \Delta \omega =0$, all the eigenvectors are orthogonal and their eigenvalues are real as well. If the eigenvectors are normalized, then
\beq
 \mathbf{B}^{(p)} \cdot \mathbf{B}^{(q)} = \delta_{pq} \,. \label{orthoB2}
\eeq 
Although the inter-bunch tune shift is not real for typical wake fields, it is for a purely reactive damper. In that case this theorem might be useful.  

This theorem is a consequence of a more general statement about orthogonality of eigenvectors of Hermitian adjoint matrices: if $\mathbf{B}$ is a set of eigenvectors of a matrix $\mathcal{M}$, and $\tilde{\mathbf{B}}$ is the same for the Hermitian adjoint matrix $\mathcal{M}^{\dagger}$, then after a proper normalization $\tilde{\mathbf{B}}^{(p)} \cdot \mathbf{B}^{(q)} =\delta_{pq}$. For Eq.(\ref{nuBk}), the matrix $\mathcal{M}_{mn}=\nu_m^0 \delta_{mn} + \delta \omega I_m I_n^*$ is self-adjoint if and only if $\Im \Delta \omega = 0$. Thus, for a real coupled-bunch tune shift $\Delta \omega$, the vectors $\mathbf{B}^{(p)}$ are orthogonal.

%-----------------------------------------------------------------%

\nocite{*}

\bibliography{bibfile}% Produces the bibliography via BibTeX.

\providecommand{\noopsort}[1]{}\providecommand{\singleletter}[1]{#1}%
\begin{thebibliography}{17}%
\makeatletter
\providecommand \@ifxundefined [1]{%
 \@ifx{#1\undefined}
}%
\providecommand \@ifnum [1]{%
 \ifnum #1\expandafter \@firstoftwo
 \else \expandafter \@secondoftwo
 \fi
}%
\providecommand \@ifx [1]{%
 \ifx #1\expandafter \@firstoftwo
 \else \expandafter \@secondoftwo
 \fi
}%
\providecommand \natexlab [1]{#1}%
\providecommand \enquote  [1]{``#1''}%
\providecommand \bibnamefont  [1]{#1}%
\providecommand \bibfnamefont [1]{#1}%
\providecommand \citenamefont [1]{#1}%
\providecommand \href@noop [0]{\@secondoftwo}%
\providecommand \href [0]{\begingroup \@sanitize@url \@href}%
\providecommand \@href[1]{\@@startlink{#1}\@@href}%
\providecommand \@@href[1]{\endgroup#1\@@endlink}%
\providecommand \@sanitize@url [0]{\catcode `\\12\catcode `\$12\catcode
  `\&12\catcode `\#12\catcode `\^12\catcode `\_12\catcode `\%12\relax}%
\providecommand \@@startlink[1]{}%
\providecommand \@@endlink[0]{}%
\providecommand \url  [0]{\begingroup\@sanitize@url \@url }%
\providecommand \@url [1]{\endgroup\@href {#1}{\urlprefix }}%
\providecommand \urlprefix  [0]{URL }%
\providecommand \Eprint [0]{\href }%
\providecommand \doibase [0]{http://dx.doi.org/}%
\providecommand \selectlanguage [0]{\@gobble}%
\providecommand \bibinfo  [0]{\@secondoftwo}%
\providecommand \bibfield  [0]{\@secondoftwo}%
\providecommand \translation [1]{[#1]}%
\providecommand \BibitemOpen [0]{}%
\providecommand \bibitemStop [0]{}%
\providecommand \bibitemNoStop [0]{.\EOS\space}%
\providecommand \EOS [0]{\spacefactor3000\relax}%
\providecommand \BibitemShut  [1]{\csname bibitem#1\endcsname}%
\let\auto@bib@innerbib\@empty
%</preamble>
\bibitem [{\citenamefont {Kourbanis}(2014)}]{Kourbanis:2014iza}%
  \BibitemOpen
  \bibfield  {author} {\bibinfo {author} {\bibfnamefont {I.}~\bibnamefont
  {Kourbanis}},\ }in\ \href
  {http://inspirehep.net/record/1315478/files/arXiv:1409.1940.pdf} {\emph
  {\bibinfo {booktitle} {{Proceedings, 5th International Particle Accelerator
  Conference (IPAC 2014)}}}}\ (\bibinfo {year} {2014})\ \Eprint
  {http://arxiv.org/abs/1409.1940} {arXiv:1409.1940 [physics.acc-ph]}
  \BibitemShut {NoStop}%
%%CITATION = ARXIV:1409.1940;%%
\bibitem [{\citenamefont {Kourbanis}(2017)}]{Kourbanis:2017zqt}%
  \BibitemOpen
  \bibfield  {author} {\bibinfo {author} {\bibfnamefont {I.}~\bibnamefont
  {Kourbanis}},\ }in\ \href {\doibase 10.18429/JACoW-NAPAC2016-THA3IO01} {\emph
  {\bibinfo {booktitle} {{Proceedings, 2nd North American Particle Accelerator
  Conference (NAPAC2016): Chicago, Illinois, USA, October 9-14, 2016}}}}\
  (\bibinfo {year} {2017})\ p.\ \bibinfo {pages} {THA3IO01}\BibitemShut
  {NoStop}%
%%CITATION = NAPAC-2016-THA3IO01;%%
\bibitem [{\citenamefont {Chao}(1993)}]{chao1993physics}%
  \BibitemOpen
  \bibfield  {author} {\bibinfo {author} {\bibfnamefont {A.~W.}\ \bibnamefont
  {Chao}},\ }\href@noop {} {\emph {\bibinfo {title} {Physics of collective beam
  instabilities in high energy accelerators}}}\ (\bibinfo  {publisher}
  {Wiley},\ \bibinfo {year} {1993})\BibitemShut {NoStop}%
\bibitem [{\citenamefont {Burov}(2009)}]{burov2009head}%
  \BibitemOpen
  \bibfield  {author} {\bibinfo {author} {\bibfnamefont {A.}~\bibnamefont
  {Burov}},\ }\href@noop {} {\bibfield  {journal} {\bibinfo  {journal}
  {Physical Review Special Topics-Accelerators and Beams}\ }\textbf {\bibinfo
  {volume} {12}},\ \bibinfo {pages} {044202} (\bibinfo {year}
  {2009})}\BibitemShut {NoStop}%
\bibitem [{\citenamefont {Balbekov}(2012)}]{PhysRevSTAB.15.054403}%
  \BibitemOpen
  \bibfield  {author} {\bibinfo {author} {\bibfnamefont {V.}~\bibnamefont
  {Balbekov}},\ }\href {\doibase 10.1103/PhysRevSTAB.15.054403} {\bibfield
  {journal} {\bibinfo  {journal} {Phys. Rev. ST Accel. Beams}\ }\textbf
  {\bibinfo {volume} {15}},\ \bibinfo {pages} {054403} (\bibinfo {year}
  {2012})}\BibitemShut {NoStop}%
\bibitem [{\citenamefont {Burov}(2013)}]{Burov:2013nb}%
  \BibitemOpen
  \bibfield  {author} {\bibinfo {author} {\bibfnamefont {A.}~\bibnamefont
  {Burov}},\ }\href@noop {} {\  (\bibinfo {year} {2013})},\ \Eprint
  {http://arxiv.org/abs/1301.1721} {arXiv:1301.1721 [physics.acc-ph]}
  \BibitemShut {NoStop}%
%%CITATION = ARXIV:1301.1721;%%
\bibitem [{\citenamefont {Burov}(2014)}]{burov2014nested}%
  \BibitemOpen
  \bibfield  {author} {\bibinfo {author} {\bibfnamefont {A.}~\bibnamefont
  {Burov}},\ }\href@noop {} {\bibfield  {journal} {\bibinfo  {journal}
  {Physical Review Accelerators and Beams}\ }\textbf {\bibinfo {volume} {17}},\
  \bibinfo {pages} {021007} (\bibinfo {year} {2014})}\BibitemShut {NoStop}%
\bibitem [{\citenamefont {Van~Kampen}(1957)}]{van1957dispersion}%
  \BibitemOpen
  \bibfield  {author} {\bibinfo {author} {\bibfnamefont {N.}~\bibnamefont
  {Van~Kampen}},\ }\href@noop {} {\bibfield  {journal} {\bibinfo  {journal}
  {Physica}\ }\textbf {\bibinfo {volume} {23}},\ \bibinfo {pages} {641}
  (\bibinfo {year} {1957})}\BibitemShut {NoStop}%
\bibitem [{\citenamefont {Burov}(2011)}]{Burov:2011zza}%
  \BibitemOpen
  \bibfield  {author} {\bibinfo {author} {\bibfnamefont {A.}~\bibnamefont
  {Burov}},\ }\bibfield  {booktitle} {\emph {\bibinfo {booktitle} {{Particle
  accelerator. Proceedings, 24th Conference, PAC'11, New York, USA, March
  28-April 1, 2011}}},\ }\href@noop {} {\bibfield  {journal} {\bibinfo
  {journal} {Conf. Proc.}\ }\textbf {\bibinfo {volume} {C110328}},\ \bibinfo
  {pages} {94} (\bibinfo {year} {2011})},\ \Eprint
  {http://arxiv.org/abs/1208.4338} {arXiv:1208.4338 [physics.acc-ph]}
  \BibitemShut {NoStop}%
%%CITATION = ARXIV:1208.4338;%%
\bibitem [{\citenamefont {Macridin}\ \emph {et~al.}(2015)\citenamefont
  {Macridin}, \citenamefont {Burov}, \citenamefont {Stern}, \citenamefont
  {Amundson},\ and\ \citenamefont {Spentzouris}}]{PhysRevSTAB.18.074401}%
  \BibitemOpen
  \bibfield  {author} {\bibinfo {author} {\bibfnamefont {A.}~\bibnamefont
  {Macridin}}, \bibinfo {author} {\bibfnamefont {A.}~\bibnamefont {Burov}},
  \bibinfo {author} {\bibfnamefont {E.}~\bibnamefont {Stern}}, \bibinfo
  {author} {\bibfnamefont {J.}~\bibnamefont {Amundson}}, \ and\ \bibinfo
  {author} {\bibfnamefont {P.}~\bibnamefont {Spentzouris}},\ }\href {\doibase
  10.1103/PhysRevSTAB.18.074401} {\bibfield  {journal} {\bibinfo  {journal}
  {Phys. Rev. ST Accel. Beams}\ }\textbf {\bibinfo {volume} {18}},\ \bibinfo
  {pages} {074401} (\bibinfo {year} {2015})}\BibitemShut {NoStop}%
\bibitem [{\citenamefont
  {Burov}(2016{\natexlab{a}})}]{PhysRevAccelBeams.19.084402}%
  \BibitemOpen
  \bibfield  {author} {\bibinfo {author} {\bibfnamefont {A.}~\bibnamefont
  {Burov}},\ }\href {\doibase 10.1103/PhysRevAccelBeams.19.084402} {\bibfield
  {journal} {\bibinfo  {journal} {Phys. Rev. Accel. Beams}\ }\textbf {\bibinfo
  {volume} {19}},\ \bibinfo {pages} {084402} (\bibinfo {year}
  {2016}{\natexlab{a}})}\BibitemShut {NoStop}%
\bibitem [{\citenamefont {Burov}(2018)}]{Burov:2018pjl}%
  \BibitemOpen
  \bibfield  {author} {\bibinfo {author} {\bibfnamefont {A.}~\bibnamefont
  {Burov}},\ }\href@noop {} {\  (\bibinfo {year} {2018})},\ \Eprint
  {http://arxiv.org/abs/1807.04887} {arXiv:1807.04887 [physics.acc-ph]}
  \BibitemShut {NoStop}%
%%CITATION = ARXIV:1807.04887;%%
\bibitem [{\citenamefont {Burov}(2016{\natexlab{b}})}]{Burov:2016jsh}%
  \BibitemOpen
  \bibfield  {author} {\bibinfo {author} {\bibfnamefont {A.}~\bibnamefont
  {Burov}},\ }\href@noop {} {\bibfield  {journal} {\bibinfo  {journal}
  {Submitted to: Phys. Rev. Accel. Beams}\ } (\bibinfo {year}
  {2016}{\natexlab{b}})},\ \Eprint {http://arxiv.org/abs/1606.07430}
  {arXiv:1606.07430 [physics.acc-ph]} \BibitemShut {NoStop}%
%%CITATION = ARXIV:1606.07430;%%
\bibitem [{\citenamefont {Eddy}\ \emph {et~al.}(2018)\citenamefont {Eddy},
  \citenamefont {Fellenz}, \citenamefont {Prieto},\ and\ \citenamefont
  {Zorzetti}}]{Eddy:2018tdm}%
  \BibitemOpen
  \bibfield  {author} {\bibinfo {author} {\bibfnamefont {N.}~\bibnamefont
  {Eddy}}, \bibinfo {author} {\bibfnamefont {B.}~\bibnamefont {Fellenz}},
  \bibinfo {author} {\bibfnamefont {P.}~\bibnamefont {Prieto}}, \ and\ \bibinfo
  {author} {\bibfnamefont {S.}~\bibnamefont {Zorzetti}},\ }in\ \href {\doibase
  10.18429/JACoW-IBIC2017-TUPCF21} {\emph {\bibinfo {booktitle} {{Proceedings,
  6th International Beam Instrumentation Conference, IBIC2017}}}}\ (\bibinfo
  {year} {2018})\ p.\ \bibinfo {pages} {TUPCF21}\BibitemShut {NoStop}%
%%CITATION = FERMILAB-CONF-17-678-AD;%%
\bibitem [{\citenamefont {Ainsworth}(2018)}]{Ainsworth:2018HB}%
  \BibitemOpen
  \bibfield  {author} {\bibinfo {author} {\bibfnamefont {e.~a.}\ \bibnamefont
  {Ainsworth}, \bibfnamefont {Robert}},\ }in\ \href@noop {} {\emph {\bibinfo
  {booktitle} {{Proceedings, High Brightness 2018 Conference}}}}\ (\bibinfo
  {year} {2018})\ p.\ \bibinfo {pages} {TUA1WD04}\BibitemShut {NoStop}%
%%CITATION = FERMILAB-CONF-17-678-AD;%%
\bibitem [{\citenamefont {Burov}(2015)}]{burov2015damping}%
  \BibitemOpen
  \bibfield  {author} {\bibinfo {author} {\bibfnamefont {A.}~\bibnamefont
  {Burov}},\ }\href@noop {} {\bibfield  {journal} {\bibinfo  {journal} {arXiv
  preprint arXiv:1505.07704}\ } (\bibinfo {year} {2015})}\BibitemShut {NoStop}%
\bibitem [{\citenamefont {Shiltsev}\ \emph {et~al.}(2017)\citenamefont
  {Shiltsev}, \citenamefont {Alexahin}, \citenamefont {Burov},\ and\
  \citenamefont {Valishev}}]{PhysRevLett.119.134802}%
  \BibitemOpen
  \bibfield  {author} {\bibinfo {author} {\bibfnamefont {V.}~\bibnamefont
  {Shiltsev}}, \bibinfo {author} {\bibfnamefont {Y.}~\bibnamefont {Alexahin}},
  \bibinfo {author} {\bibfnamefont {A.}~\bibnamefont {Burov}}, \ and\ \bibinfo
  {author} {\bibfnamefont {A.}~\bibnamefont {Valishev}},\ }\href {\doibase
  10.1103/PhysRevLett.119.134802} {\bibfield  {journal} {\bibinfo  {journal}
  {Phys. Rev. Lett.}\ }\textbf {\bibinfo {volume} {119}},\ \bibinfo {pages}
  {134802} (\bibinfo {year} {2017})}\BibitemShut {NoStop}%
\end{thebibliography}%

\end{document}